\documentclass[11pt]{article}
\usepackage{moriond,epsfig}

\def\ra{\rightarrow}

\def\be{\begin{equation}}
\def\ee{\end{equation}}
\def\bea{\begin{eqnarray}}
\def\eea{\end{eqnarray}}

\def\Zzero{{\rm Z}^0}
\def\epem{e^{+}e^{-}}
\def\ccbar{c\overline{c}}
\def\bbbar{b\overline{b}}
\def\uubar{u\overline{u}}
\def\ddbar{d\overline{d}}
\def\ssbar{s\overline{s}}
\def\qqbar{q\overline{q}}
\def\qqbarg{q\overline{q}g}
\begin{document}
\vspace*{4cm}
\title{HADRONISATION AT LEP}

\author{ ELI BEN-HAIM }
\address{Laboratoire de l'Acc\'el\'erateur Lin\'eaire (L.A.L.),\\ 
Universit\'e Paris-Sud, B\^atiment 200,\\
BP 34, F-91898 Orsay cedex, France\\[10pt]
\epsfig{file=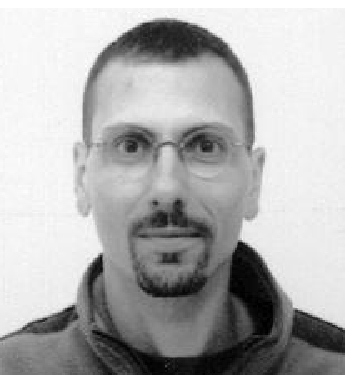,width=35mm}}

\maketitle\abstracts{
An overview of recent results from LEP concerning the hadronisation process 
is presented. Emphasis is placed on the $b$-quark. The first presented 
analysis is the measurement of the $b$-quark fragmentation function. It 
includes a new, hadronic-model independent method to extract the x-dependence 
of the non-perturbative QCD component from the measured fragmentation function.
 This is followed by the results of two analyses on, respectively,
production rates of $b$-excited states and branching fractions of $b$-quarks 
to neutral and charged $b$-hadrons. Multiplicity in the final state is also 
discussed concerning the difference in multiplicities between $b$ and light 
quark initiated events, and total multiplicities in three jet events. 
Finally, recent measurements of $\omega$ and $\eta$ meson production rates 
are given.}

\section{$b$-Quark Fragmentation Function}
\subsection{LEP Measurements}\label{subsec:fraglep}

\begin{figure}[t]
\begin{minipage}[t][1cm][t]{7.cm}
\includegraphics[scale=0.35]{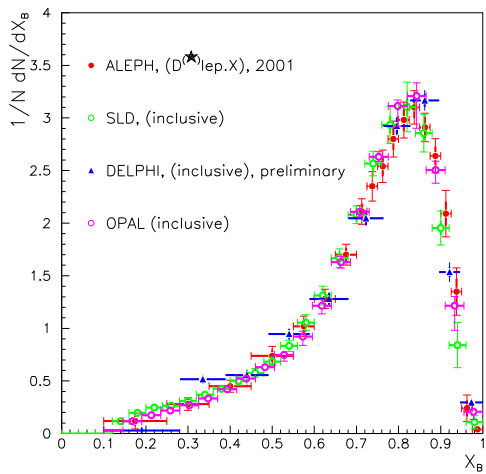}
\parbox{6.5cm}{\caption{Unfolded x distributions from ALEPH, DELPHI, OPAL and SLD.} 
\label{fig:fragdist}}
\end{minipage} 
\begin{minipage} [t][2cm][t]{8.9cm} 
\vspace{-6.5cm}
component (PC) describing gluon emission by the quarks and a non-perturbative QCD component (NPC) which incorporates all mechanisms at work to bridge  the gap between the previous phase and the production of weakly decaying $b$-mesons. The PC can be obtained using analytic expressions or Monte Carlo generators. The NPC is usually parametrised phenomenologically via a model. A method is  proposed to extract the NPC directly, and independently  of any hadronic physics modelling~\cite{notre}. The extraction has been done using some of the measurements of the $b$-quark fragmentation function mentioned in Section~\ref{subsec:fraglep}. The method employs direct and inverse Mellin transformation. Its detailed description is outside the framework of this paper, and therefore we would concentrate on results. The extracted NPC depends only on the way the PC has been defined.
\end{minipage}
\vspace{-1.3cm} 
\end{figure}

The $b$-quark fragmentation function in $\epem$ collisions is commonly 
defined as the distribution of the scaled energy $x=\frac{E_B}{E_{beam}}$
variable, 
where $E_B$ is the energy taken by the weekly decaying $b$-hadron.
New measurements of the fragmentation function at or near the $\Zzero$ pole are 
now available from ALEPH~\cite{alephfrag}, DELPHI (preliminary)~\cite{delphifrag}, 
OPAL~\cite{opalfrag}, and SLD~\cite{sldfrag}. These results are presented 
in Figure~\ref{fig:fragdist}. The different collaborations chose different methods 
for the reconstruction of the b-hadron's energy and for the unfolding of the 
underlying $x$ from the measured one. This measurement is affected by the 
control of the finite resolution of the detector, which also gives the main 
systematic uncertainty.

\subsection{Extraction of the Non-Perturbative QCD Component}\label{subsec:extract}

\begin{figure}[ht]
\begin{center}
\mbox{\epsfig{file=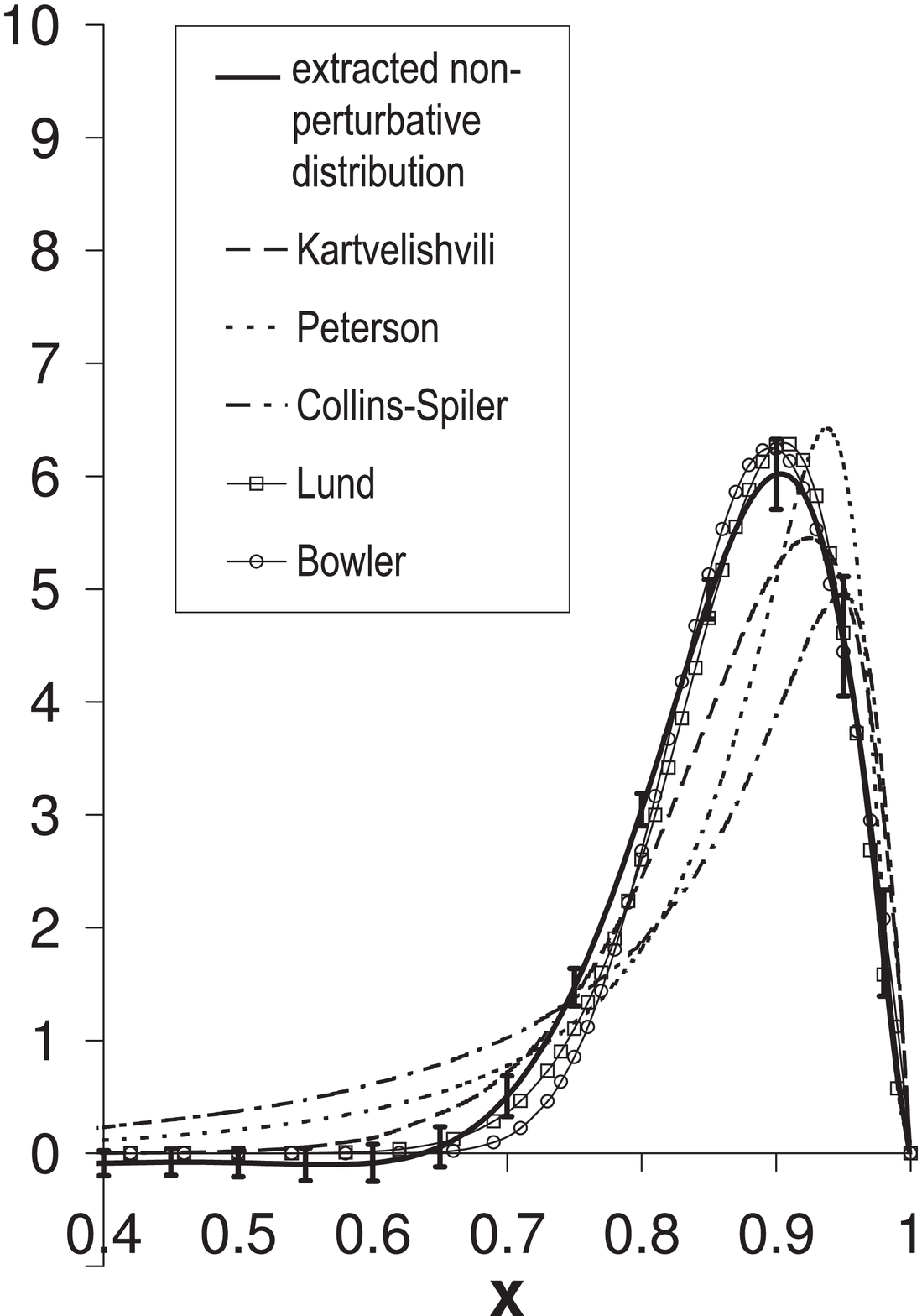,width=7cm,height=7cm}
      \epsfig{file=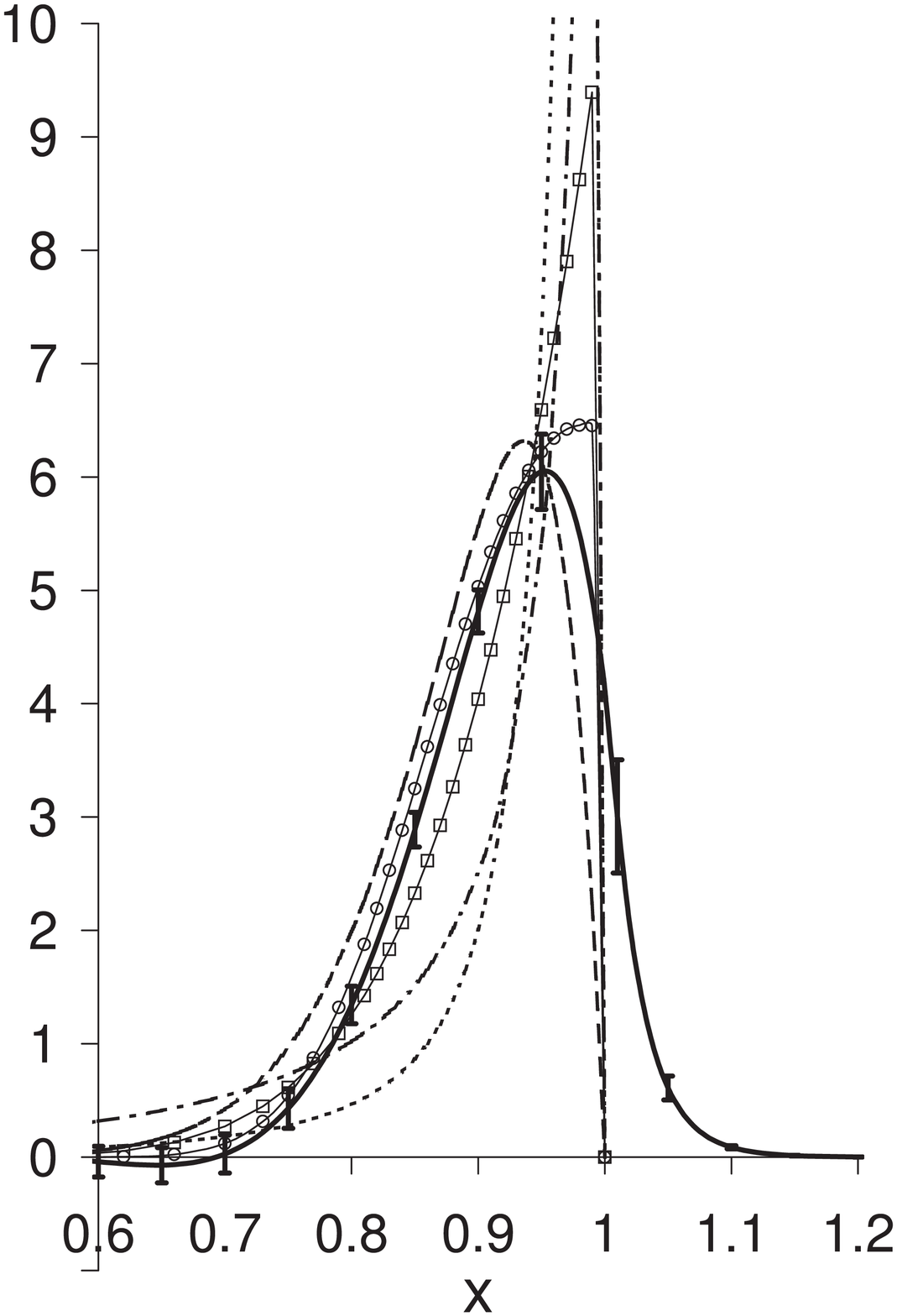,width=7cm,height=7cm}}
\end{center}
\caption[]{Comparison between the directly extracted non-perturbative 
component (thick full line) and the model fits on ALEPH's data. 
Left: the perturbative QCD component is taken from JETSET. 
Right: the theoretical perturbative QCD component is used.
\label{fig:notre}}
\end{figure}

The $b$-quark fragmentation function is generally viewed as resulting from three components: the primary interaction ($\epem$ annihilation into  a $b \overline{b}$ pair in the present study), a perturbative QCD 
The method has been applied using two different approaches to evaluate the PC: a parton shower Monte-Carlo, and  an analytic NLL computation~\cite{cacciari}. In both cases, the extracted functions have been found to be zero for $x < 0.7$. This means that the gluon radiation is well accounted, in this region, by both approaches. Results for the extracted NPC and  comparisons with distributions from known hadronisation models, that have been fitted to the ALEPH data~\cite{alephfrag}, are shown in Figure~\ref{fig:notre}. When the PC is taken from a parton shower Monte-Carlo, the NPC is rather similar in shape with those obtained from the Lund or Bowler models, and is rather different from distributions obtained in other models that have tails at low x (e.g. Peterson, Collins-Spiller). When the PC is the result of an analytic NLL computation, the NPC has to be extended beyond $x = 1$. This has not a physical meaning, It is directly related to the breakdown of the NLL QCD approach when $x$ gets close to $1$, and is necessary in order to compensate for the unphysical behaviour of the PC in this region. The fact that the NPC extends into a non-physical region implies that it cannot be described by any  hadronic modelling, in this situation. In the two perturbative approaches that have been studied, it happens that the NPC has a similar shape, being simply translated to higher-$x$ values in the case of analytic NLL QCD, illustrating the ability of this approach to account for softer gluon radiation than with a parton shower generator. The NPC extracted in the proposed way can then be used in another environment than $\epem$ annihilation, as long as the same parameters and methods are taken for the evaluation of the PC. Consistency checks, on the matching between the measured and predicted $b$-fragmentation distribution, can be defined which provide information on the determination of the PC itself.

\section{Excited $b$-Hadron States}

\begin{figure}[htb]
\begin{center}
\mbox{\epsfig{file=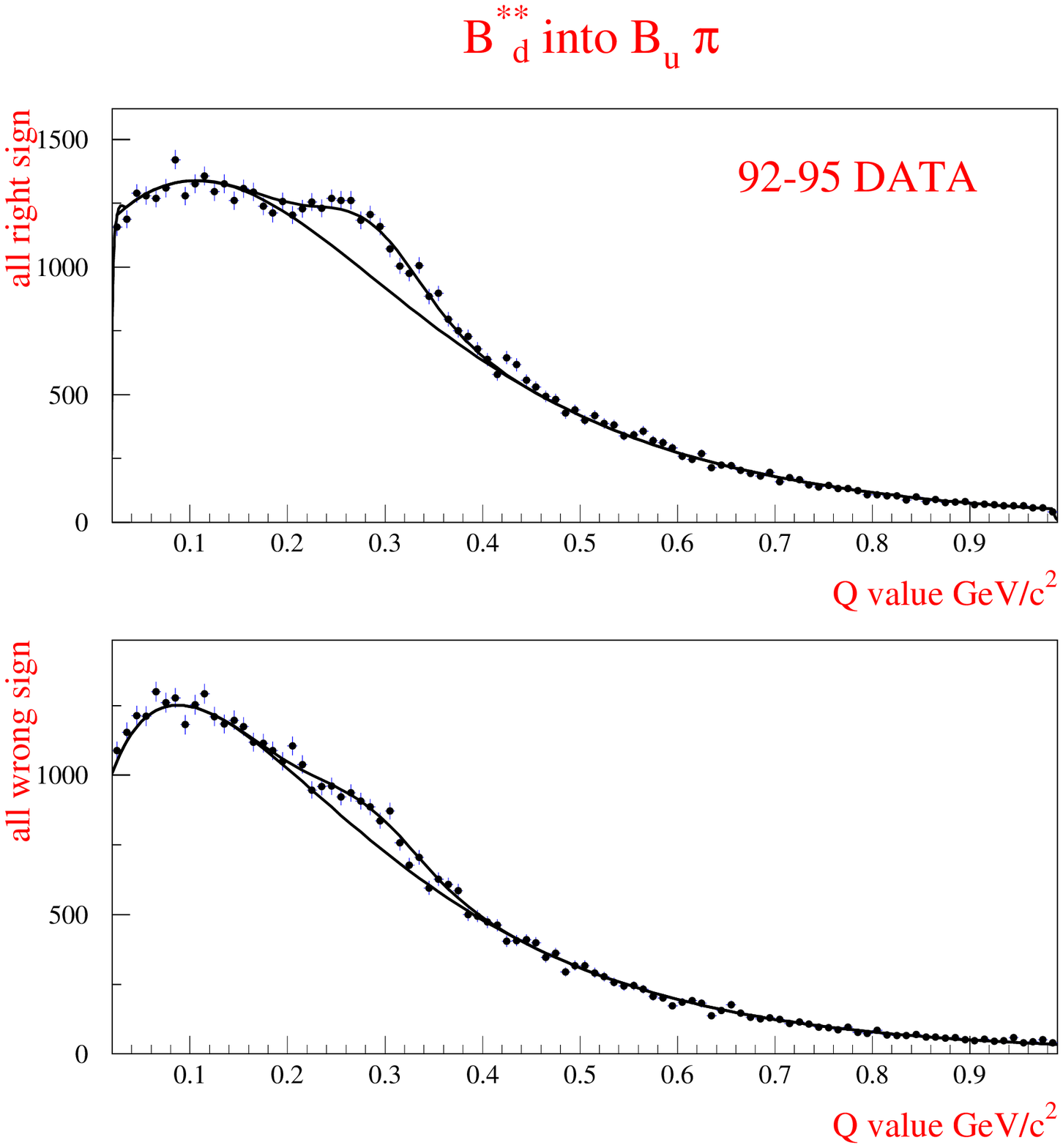,width=7cm,height=7cm}
      \epsfig{file=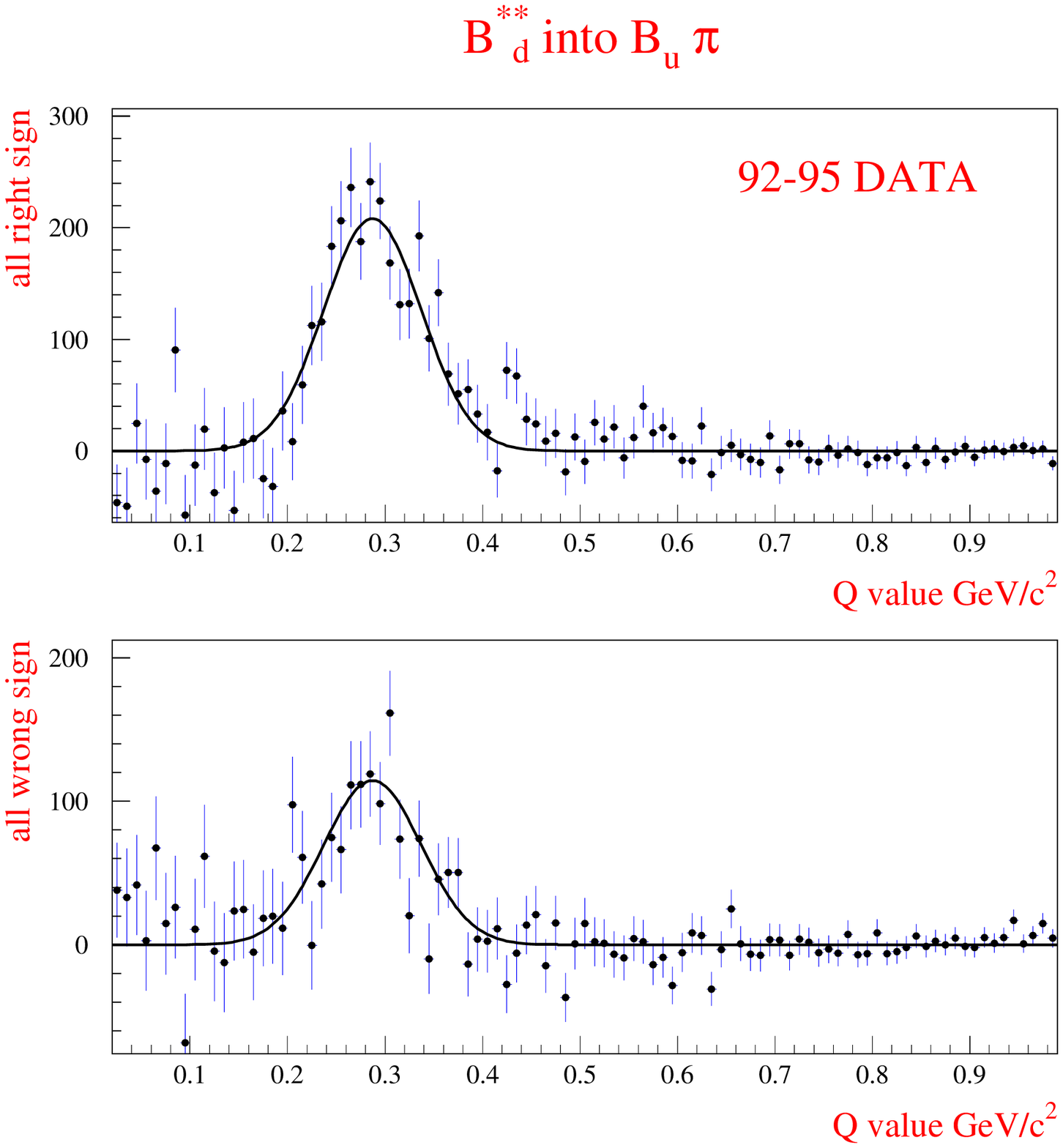,width=7cm,height=7cm}}
\end{center}
\caption{The $Q$ distributions for the $\overline{B^{**}_d}$ signal enriched (top) and background enriched (bottom) samples. The result of the fit is superimposed. On the right, the $Q$ distributions after subtraction of the fit result for the background.
\label{fig:ex}}
\end{figure}

A new analysis from DELPHI concerning the spectroscopy of excited $B$ states from 
the LEP data set of the years 1992-98 is reported~\cite{delphiex}. Results are still
preliminary. In this analysis the distribution of the variable 
$Q=m(B^{(*)}\pi)-m(B^{(*)})-m(\pi)$ is considered. The background shape is extracted 
from data, and therefore the dependence on Monte Carlo is strongly reduced.
The method is based on the use of two data samples
which are, respectively, enriched and depleted in signal events.
The background shape is 
determined directly from data.
The two data samples are fitted simultaneously by a sum of two functions: one 
for signal and the other for background (Figure~\ref{fig:ex}). The production rate 
of narrow $B^{**}_{u,d}$ states - assuming that they can be described by a single 
Gaussian distribution- is measured to be: 
\begin{equation}
P_{\bar{b}}(B^{**}_u)_{narrow} + P_{\bar{b}}(B^{**}_d)_{narrow} = 0.098 \pm 0.007 \pm 0.012.
\end{equation}
Data suggest also the presence of broad states. The new measured 
rate, for narrow states, differs markedly from the one of $\sim 20\%$ usually 
reported. Nevertheless it is compatible with the measured production rate for 
narrow $D^{**}$ states in $c$-jets. 
For $B^{**}_s$ and $\Sigma_b^{(*)}$ upper limits on the production rates are 
obtained: $P_{\bar{b}}(B^{**}_s) < 0.015$ , $ P_b(\Sigma_b^{(*)})< 0.015$, at $95\%$ 
confidence level, which supersede previous DELPHI results on these states.

\section{Branching Fractions of the b-Quark into Neutral and Charged b-Hadrons}

The production fractions of charged and neutral weakly decaying $b$-hadrons 
in $b$-quark events have been measured with the DELPHI 
detector~\cite{delphibf}. An algorithm has been developed, based on a neural 
network, to estimate the charge of the weakly decaying $b$-hadron by 
distinguishing decay particles from their fragmentation counterparts. 
From the data taken in years 1994 and 1995, the fraction of positively 
charged $b$-hadrons has been measured to be: 
$f^+ = (42.06 \pm 0.81(stat.) \pm 0.91(syst.))\%$. Subtracting the rates for 
charged $\bar{\Xi}_b^+$ and $\bar{\Omega}_b^+$ baryons gives the production 
fraction of $B^+$ mesons: 
$f_{B_u} = (40.96 \pm 0.81(stat.) \pm 1.14(syst.))\%$. 
This is at present the most accurate measurement available.
 
\section{Charged Particle Multiplicities}

\begin{figure}[htb]
\begin{center}
\includegraphics[scale=0.7]{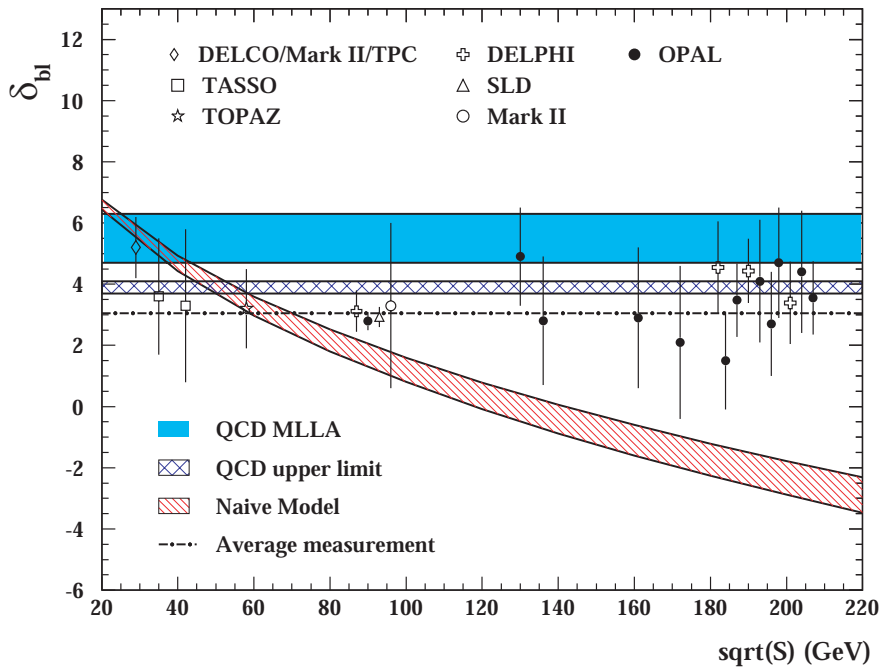}
\end{center}
\caption{$\delta_{bl}$ as a function of the centre-of-mass energy from 
DELPHI, OPAL and lower energy experiments. The original MLLA prediction
is shown as a shaded area to include the errors of experimental origin on this
prediction, not including missing higher order corrections. The crosshatched 
area corresponds to the QCD upper limits.
The single hatched area represents the naive model prediction.
\label{fig:mult}}
\end{figure}

The mean charged particle multiplicities have been measured separately for 
$\bbbar$, $\ccbar$ and light quark ($\uubar$, $\ddbar$, $\ssbar$) initiated 
events produced in $\epem$ annihilations at LEP. The data were recorded with 
the DELPHI~\cite{delphimult} and OPAL~\cite{opalmult} detectors at energies 
above the $\Zzero$ peak, up to 206 GeV, corresponding to the full statistics 
collected at LEP1.5 and LEP2. The difference in mean 
charged particle multiplicities for $b$ and light quark events, $\delta_{bl}$, 
measured over this energy range is consistent with an energy independent 
behaviour, as predicted by a QCD calculation within the Modifed Leading Log 
Approximation (MLLA)~\cite{mlla}, based on coherence of gluon radiation. 
It is inconsistent with the prediction of 
a more phenomenological approach, the naive model~\cite{naive}, 
assuming the additional multiplicity accompanying the haevy quark to be
emitted incoherently. Lower energy measurements could not 
discriminate between the two approaches. The combined OPAL and DELPHI results 
yield at average energies close to 196 GeV 
$\delta_{bl} = 3.44 \pm 0.40 (stat.) \pm 0.89 (syst.)$ and 
$\delta_{bl} = 4.26 \pm 0.51 (stat.) \pm 0.46 (syst.)$, respectively. These 
measurements, together with those from lower energy experiments, are presented 
in Figure~\ref{fig:mult}.

\section{Multiplicities in 3 Jet events}

Data collected by the DELPHI detector have been used to determine the charged 
hadron multiplicity in three jet events ($\qqbarg$). The multiplicity has been 
measured in a cone of $30^{\circ}$ opening angle perpendicular to the event 
plane. This cone multiplicity is noted $N_{ch}(30^{\circ})$. In~\cite{khose} 
a prediction is made, based on the assumption of coherent gluon radiation, 
relating the multiplicity $N_{\bot}^{\qqbarg}$ in cones perpendicular to the 
event plane in 3 jet events to the multiplicity $N_{\bot}^{\qqbar}$ in cones 
perpendicular to the event axis in 2 jet events:

\begin{equation}
N_{\bot}^{\qqbarg} = r \cdot N_{\bot}^{\qqbar}
\label{eq:Nperp}
\end{equation}
The function $r$ depends on the colour factors and the interjet angles 
( $\theta_{qg}$, $\theta_{\overline{q}g}$, $\theta_{\qqbar}$):

\begin{equation}
r=\frac{C_A}{4C_F}
\left [
(1-\cos \theta_{qg}) + (1-\cos \theta_{\overline{q}g}) - \frac{1}{C_A^2}(1-\cos \theta_{\qqbar})
\right ]
\label{eq:momdata}
\end{equation}

\begin{figure}[htb]
\begin{center}
\includegraphics[scale=0.7]{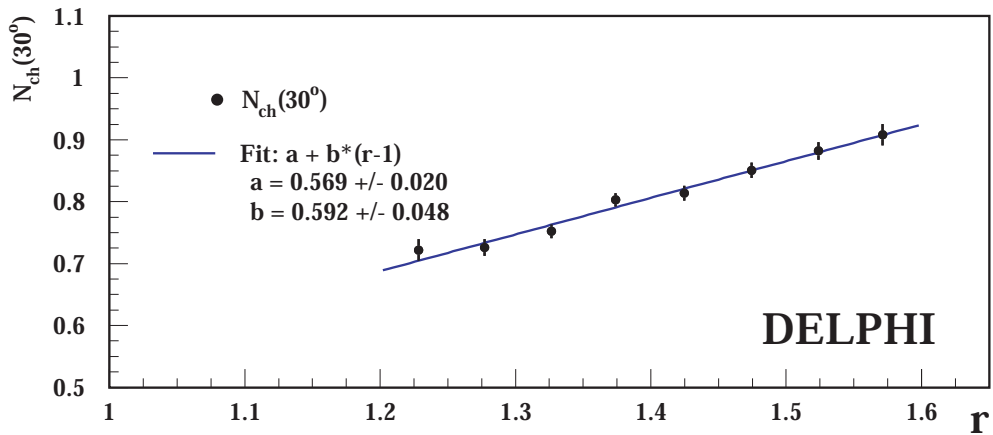}
\end{center}
\caption{$N_{ch}(30^{\circ})$ as a function of $r$ measured by DELPHI and a straight line fit (preliminary).
\label{fig:3jet}}
\end{figure}

The measured points are shown in Figure~\ref{fig:3jet}. They have been fitted 
using a linear function of the form: $a+b(r-1)$. The values for $a$ and $b$ 
are 
also shown on the plot. They are compatible with each other and with the 
previously measured value $N_\bot^{\qqbar}(30^{\circ})=0.593\pm0.001$.
Therefore they support the first order prediction of Equation~\ref{eq:momdata} based 
on the coherence of gluon radiation.

\section{Production Rates of $\omega$ and $\eta$ Mesons}

A new measurement of the inclusive production of the $\omega(782)$ and $\eta$ 
mesons in hadronic $\Zzero$ is available from ALEPH~\cite{alephomega}. The 
analysis is based on 4 million hadronic $\Zzero$ decays recorded between 1991 
and 1995. The production rate for $x_p = p_{meson}/p_{beam} > 0.05$ is 
measured in the $\omega \ra \pi^+\pi^-\pi^0$ decay mode and found to be 
$0.585 \pm 0.019(stat.) \pm 0.033 (syst.) $ per event. Inclusive $\eta$ meson 
production is measured in the same decay channel for $x_p > 0.10$, obtaining 
$0.355 \pm 0.011 (stat.) \pm 0.024 (syst.)$ per event. These results are 
compatible with previous observations from L3~\cite{L3omega} and 
OPAL~\cite{opalomega}.

\section*{Acknowledgments}
I would like to thank all physicists from LEP experiments who provided me with 
their last results for this presentation. My work is supported by EEC RTN contract 
HPRN-CT-00292-2002.

\end{document}